\documentstyle[12pt]{JHEP3}

\title{Plane Wave Type II$^*$ String Backgrounds}

\author{Harvendra Singh\footnote{hsingh@iitg.ernet.in} \\
Department of Physics, \\
Indian Institute of Technology, Guwahati-781039, Assam, India}

\abstract{ In this note
we aim to study plane-wave limits of the solutions  of
type II$^*$ superstring theories. We consider  Freund-Rubin type
$dS_5\times H^5$ solutions of type IIB$^*$ theory
and  obtain a new kind of plane-wave solutions, we refer them as de
Sitter plane-waves or
Dpp-waves. Considering Hull's time-like T-duality we are able to map
the Dpp wave solution to maximally supersymmetric Hpp-wave in IIB string
theory and vice-versa.}
\preprint{hep-th/0405193}
\preprint{ \today}
\begin{document}
\def\be{\begin{equation}} \def\ee{\end{equation}}
\def\bea{\begin{eqnarray}} \def\eea{\end{eqnarray}} \def\ba{\begin{array}}
\def\ea{\end{array}} \def\ben{\begin{enumerate}} \def\een{\end{enumerate}}
\def\nab{\bigtriangledown} \def\tpi{\tilde\Phi} \def\nnu{\nonumber}
\def\lll{\label}
\newcommand{\eqn}[1]{(\ref{#1})}
\def\cC{{\cal C}}
\def\cG{{\cal G}}
\def\cd{{\cal D}}
\def\a{\alpha}
\def\b{\beta}
\def\g{\gamma}\def\G{\Gamma}
\def\d{\delta}\def\D{\Delta}
\def\ep{\epsilon}
\def\e{\eta}
\def\z{\zeta}
\def\t{\theta}\def\T{\Theta}
\def\l{\lambda}\def\L{\Lambda}
\def\m{\mu}
\def\f{\phi}\def\F{\Phi}
\def\n{\nu}
\def\p{\psi}\def\P{\Psi}
\def\r{\rho}
\def\s{\sigma}\def\S{\Sigma}
\def\ta{\tau}
\def\x{\chi}
\def\o{\omega}\def\O{\Omega}
\def\k{\kappa}
\def\pa {\partial}
\def\ov{\over}
\def\br{\nonumber\\}
\def\ud{\underline}
\section{Introduction}
The supersymmetric plane-wave (pp-wave)
backgrounds in string theory \cite{blau,blau1}
could, generically, be obtained by applying the Penrose-like
limits \cite{penrose,gueven,blau1} on anti de Sitter spacetimes,
$AdS \times Sphere$. Under these
 plane-wave limits
\cite{blau1}  we basically zoom in onto those null geodesics
which have a direction along the sphere. If the null geodesic is
chosen such
that it does not have a component along  sphere then we obtain
flat-space solutions.
Although,  the
pp-waves are (asymptotically) non-flat geometries,
nevertheless string theory in these
backgrounds becomes exactly solvable theory in
suitable light-cone gauge \cite{matsaev,matsaev1}.
Therefore, from AdS/CFT correspondence point of view \cite{maldacena},
a pp-wave
spacetime in bulk has useful consequences for
dual conformal field theories on the boundary \cite{bmn}.
Under  BMN-correspondence \cite{bmn} the conformal field theory
operators with
large $U(1)$ R-charge are dual to type IIB closed string excitations
in a pp-wave background spacetime.
In this work we would like to concentrate on the plane-wave geometries
in type II$^*$ string theories \cite{hull}.

It was proposed some time back by Hull \cite{hull} that if we consider
time-like
T-duality
of usual type II string
theory with spacetime signature $(1,9)$, then one could  study string
theories
on arbitrary $(m,n)$ signature spacetimes as well. This duality along
time direction some how
takes
away the significance of the spacetime signature from the string theory, in
particular the significance of Lorentzian symmetry  $SO(1,9)$.
However, while implementing time-like duality it requires the assumption
of a compact
time coordinate. It also makes the RR-potential fields of the dual theory
tachyonic (ghost-like) in nature, that is, the kinetic terms with negative
sign  appear in the low energy effective action. Such type II string
theories with
tachyonic kinetic terms have been designated as II$^*$ string theories
\cite{hull}.
Nevertheless, these star-theories are maximally supersymmetric and
hopefully the quantum
fluctuations cancel out in the spectrum which could bring stability to
the flat spacetime which is still the vacuum solution.
We note that the theories with tachyonic matter fields or with a tachyonic
potential are  favourable
candidates for cosmology,
see \cite{ashoke} and  references therein, so that we are able to obtain
de-Sitter spacetime. The de Sitter spacetimes naturally appear as
solutions in
II$^*$  theories. The
 usual string theory forbid de-Sitter space as a classical solutions,
but see \cite{kachru}
for recent developments.

This paper is organised as follows. We first discuss the Penrose
limits in the case of type II$^*$ supergravity in section-2.
Then we consider
de Sitter solution of type IIB$^*$
supergravity and apply plane-wave limits to obtain the new kind of
de Sitter plane-wave (Dpp) background which can exist only in type II$^*$
string theory.
In section-3 we discuss time-like T-duality and the relationship between
Hpp and Dpp waves.
The conclusions are given in
section-4.

\section{A case for de-Sitter plane-wave (Dpp-wave)}
\subsection{Review: The Penrose Limits}
The plane-wave limits  of
anti-de Sitter spacetime, $
AdS_p\times S^q $, along a null geodesic with a
 non-zero component along the sphere leads to
a plane-wave geometry. The procedure is well described in
\cite{blau1} where maximally supersymmetric Hpp-wave solutions are
obtained by taking the limits of $AdS_5\times S^5$ type IIB
background. Our interest here is to study plane-wave limits for the
solutions of type II$^*$ superstring theories for which there exist
de Sitter solutions.

In type II$^*$ supergravity \cite{hull}, the  action has the kinetic terms
for all the RR-potentials with wrong sign, so they are tachyonic.
Originally these
theories are obtained by
implementing T-duality along compact time-like directions of ordinary
type II string theory. Assuming a compact
time like direction in a theory is pathological, but
after the transformations these tachyonic string theories have usual
Minkowskii
signature (1,9) and the maximal supersymmetry.\footnote{ The
assumption of a compact time direction is  considered only as a tool
to implement time-like duality \cite{hull}.
However, the RR-sector kinetic energy terms having minus sign
is somewhat troublesome. But it is nevertheless
hoped that the supersymmetry takes care of these instabilities since the
dual string theory is well defined.}
  It is straightforward and obvious that
II$^*$-string theories will  have following
  scaling limits on the  fields
\bea\label{pen}
  g_{\m\n}\to \xi^{-2} g_{\m\n},~~ \f\to\f ,~~
 A_{(p)} \to \xi^{-p} A_{(p)}.
\eea
 under which the action scales homogeneously. The
 scale parameter  $\xi$ has to be  strictly positive.
\footnote{ We remark that such scaling symmetry of action
could be used to tune the mass scales, if any, in a  theory like
in massive/gauged type IIA, see \cite{har03}.}

The  Penrose limits are  of the usual type
\bea\lll{pen1}
 && \bar g_{\m\n}=\lim\limits_{\O\to0} \O^{-2} g_{\m\n}(\O) \br
&&\bar\f =\lim\limits_{\O\to0} \f(\O)\br
&& \bar A_{(p)} =\lim\limits_{\O\to0} \O^{-p} A_{(p)}(\O)
\eea
where parameter $\O$ is positive.
Now the proposal is that if there exists a (anti) de-Sitter solution with
the local data
$((A)dS, g, \f, A_p)$, by implementing the limits \eqn{pen1} we will get
to new data $(PP, \bar g, \bar\f,\bar A_p)$ which is a Hpp
(Dpp)-wave
 of type IIB (IIB$^*$) string theory.
 In the next secion we will explicitly show this  by taking
an example of de-Sitter background of IIB$^*$ supergravity, the other case
of Hpp-wave is by now well studied \cite{blau1}.
It is worth noting that the limits \eqn{pen1} are the
same as in Blau et al \cite{blau1}.

\subsection{The $dS_5\times H^5$ and a Dpp-wave solution}

The type IIB$^*$ supergravity \cite{hull} the action has the 
kinetic terms of all the
RR-potentials with wrong sign, see Appendix. 
We shall assume that the IIB$^*$  theory
has a stable vacuum and is a consistent string theory. Nevertheless
there exists  a de-Sitter solution $dS_5\times H^5$ \cite{liu,hull1}.
 This de Sitter solution is given
by
\bea\lll{ss21}
&&g:= l^2 \big\{ -dt^2 + cosh^2t \left({dr^2\over 1-r^2} +r^2
  d\O_3^2\right)\big\} + l^2\big\{ d\psi^2+cosh^2\p \left({ds^2\over
1+s^2} +s^2
  d\O_3^2\right)\big\} \br
&& F_{(5)}= {2\sqrt{2}~ l^4}({\rm Vol}(dS_5)+ {\rm Vol}(H_5)),
 \eea
where
$l$ represents
the radii of the de-Sitter
$dS_5$ and hyperbolic space $H^5$, and $d\O_n^2$ is the
line element of unit $n$-sphere.
The volume 5-forms are defined for the unit size five-spaces, e.g.,
$$ {\rm Vol}(dS_5) =  cosh^4t{r^3\ov\sqrt{1-r^2}} dtdr\o_3~,$$
where $\o_3$ is the volume form over unit radius 3-sphere.
The  five-form field strength in \eqn{ss21} is such that it is
self-dual.
The rest of the background fields like dilaton etc. are vanishing in
\eqn{ss21}.

As usual in the case of plane-wave limits, we would like to change the
coordinates in the $(\p,t)$ plane to
\be
U=\p +  t ,~~~~V=\p -  t,
\ee
in terms of which the background becomes,
  \bea\lll{s21}
&&l^{-2} g:= dUdV +  cosh^2({U-V\over2})
  \left({dr^2\over 1-r^2} +r^2  d\O_3^2\right) +
cosh^2({U+V\ov2}) \left({ds^2\over 1+s^2} +s^2  d\O_3^2\right)\br
&& l^{-4} F_{(5)}={2\sqrt{2}} \{{\rm Vol}({dS_5(U,V)}) +
{\rm Vol}({H_5(U,V)})\},
 \eea
 We shall now rescale the coordinates as
\bea
U=u,~~~V={v\ov (l)^2},~~~
Y^a={y^a\ov l},~~~
Y^\a={y^\a\ov l}
\eea
where $r^2=Y^aY^a,~s^2=Y^\a Y^\a$.
Then we take the  limits \eqn{pen1},  $l\to\infty$, i.e.
large radius limit, along the null geodesic
parametrised by $U$.\footnote{ These limits involve
zooming in onto a null geodesic with component along $H^5$, the
 limits  basically
are the  Penrose limits in \cite{blau1}.}
This consists in dropping the
dependence on the coordinates other than $u$. In this way we get
the pp-wave solution  written in Rosen coordinates and
depending only on $u$,
\bea
&&\bar g:= dudv +  cosh^2 ({u\over2}) \sum_{a=1}^{4}
(dy^a)^2+cosh^2({u\ov2})\sum_{\a=5}^{8}(dy^\a)^2 \br
&&  \bar F_{(5)}= 2\sqrt{2}~\{
cosh^3({u\ov2})~dudy^1dy^2dy^3dy^4+
cosh^3({u\ov2})~dudy^5dy^6dy^7dy^8\}.
\eea
In order to write the above background in the familiar form, we shall like 
to switch to the  Brinkman coordinates $(dx^+,
dx^{-},x^a)$ as in \cite{blau}, but these transformations are slightly
different.
It is not difficult to guess that the new coordinate
relations are
\bea
x^{-}=u/2,~~~x^{+}=v -{1\ov4} \sum_{i=1}^8 {sinh(2\l_i u)\over2\l_i} y^i
y^i, ~~~
x^i= y^i {cosh(\l_iu)\over2\l_i}.
\lll{3a}\eea
After some straightforward steps
we obtain the Dpp-wave metric  in Cahen-Wallach
spacetime form
\bea\lll{newwave}
&& g:= 2dx^{+}dx^{-} +(A_{ab}x^ax^b)  (dx^{-})^2+
\sum_{a=1}^{8}(dx^a)^2\br
&& F_{-1234}=2\sqrt{2} =F_{-5678}
\eea
with the matrix
\be
A_{ab}\equiv  \delta_{ab}.
\lll{matr}
\ee
It can be calculated that the only nonvanishing
componant of the Ricci tensor for the Dpp-wave metric \eqn{newwave} is
$$R_{--}=-8  \ , $$
which is strictly negative.\footnote{We follow the convention where
for a de
Sitter space of radius of curvature $l$, the Ricci tensor is given by
$R_{\m\n}={(D-1)\over l^2}g_{\m\n}$. Here $D$ is the spacetime
dimension.} Also note that the
matrix $A$ in \eqn{matr} is diagonal and positive  which is opposite of
the case of Hpp-wave \cite{blau}.
It is obvious because the background \eqn{newwave} satisfies
the field equations
of  type IIB$^*$ supergravity where $C_{(4)}$ kinetic term has opposite
sign unlike in ordinary IIB theory.

\section{ The time-like T-duality}
\subsection{Dpp-wave from Hpp-wave}
The time-like T-duality
map between the fields of two types of strings has been constructed by
Hull
\cite{hull}. Under this map the background fields of type IIB (IIA) string
theory
are mapped to the fields of IIA$^*$ (IIB$^*$) theory and vice-versa.
Consider the
Hpp-wave
\cite{blau}
\bea\lll{Hpp}
&& g:= 2dx^{+}dx^{-} +(H_{ab}x^ax^b)  (dx^{+})^2+
\sum_{a=1}^{8}(dx^a)^2\br
&& F_{(5)}= 2\sqrt{2}\mu~ dx^{+}(
dx^1dx^2dx^3dx^4
+ dx^5dx^6dx^7dx^8)
\eea
with the matrix $H_{ab}=-
\mu^2\delta_{ab}$. The Ricci tensor for Hpp-wave is
$$R_{--}=8\mu^2  \ , $$
which is positive  unlike the Dpp-wave. In \eqn{Hpp} we can consider
$x^{+}$ as time coordinate while $x^{-}$ acts like a
space-like direction. So the 5-form flux is electric type. We will take
$x^{+}$ coordinate of \eqn{Hpp} and make
a time-like duality \cite{hull}. Using the duality relations given in the
Appendix, the background we obtain as a result is
IIA$^*$ extended string solution
\bea\lll{funstr}
&& g:= W^{-1}\left\{-(dx^{+})^2+ (dx^{-})^2\right\}+
\sum_{a=1}^{8}(dx^a)^2\br
&& F_{(4)}= 2\sqrt{2}\mu~ (dx^1dx^2dx^3dx^4 +dx^5dx^6dx^7dx^8)\br
&& B_{(2)}= -W^{-1} dx^{+}dx^{-},~~e^{2\f}=W^{-1}
\eea
where $B$ is the NS-NS field under which fundamental strings are charged
with and the function  $W=\mu^2 \delta_{ab}x^ax^b$.
$F_{(4)}$ is the tachyonic R-R four-form field strength. It is interesting
at the first place to note that such a nontrivial string-like solution, in
presence of tachyonic RR-fluxes, exists
for IIA$^*$ theory action \eqn{IIa}. We, though, make it clear  that
an extended fundamental string solution \cite{atish} when tachyonic RR
backgrounds are vanishing is anyway a solution of II$^*$ theories.

In the next step we make a space-like T-duality by compactifying the
space like direction $x^{-}$ of the background \eqn{funstr}. This 
gives us
a type IIB$^*$ solution
\bea\lll{newwave1}
&& g:= 2dx^{+}dx^{-} +(\mu^2\delta_{ab}x^ax^b)  (dx^{-})^2+
\sum_{a=1}^{8}(dx^a)^2\br
&& F_{(5)}=2\sqrt{2}\mu dx^{-}
(dx^1dx^2dx^3dx^4+dx^5dx^6dx^7dx^8),
\eea
which is nothing but the Dpp-wave given
in \eqn{newwave} when $\mu=1$. In eq.\eqn{newwave1} the coordinate
$x^{-}$ behaves like a space direction, the flux $F_{(5)}$ is magnetic
type. Hence the roles of $x^{+}$ and $x^{-}$ have got exhanged under the
above duality
map which includes one time-like T-duality.

Thus we see that starting from Hpp-wave of IIB string and implementing a
time-like duality and a space-like duality in succession, we have obtained
Dpp-wave which is a
solution of the IIB* string theory. This is confirmation of the fact that
the Hull's time-like duality map works for the case of plane-waves.

The amount of supersymmetries preserved by
the Dpp-wave solution  \eqn{newwave} is not clear as it is not straight
forward to work out the Killing spinors due to missing ingredients
like supersymmetry variations of the II$^*$ theories. But as we know type
II$^*$ string theories have 32 supersymmetries \cite{hull}.
So we would like to claim that the supersymmetry
preserved by the Dpp-wave solutions is  the same as those in
Hpp-wave,
which are maximally supersymmetric plane-wave solutions of  IIB
strings. The main reason behind this proposal is that
the Hpp and Dpp are time-like T-dual backgrounds as we have seen above.
If the time-like T-duality preserves the supersymmetry of the background
then Dpp-waves are maximally supersymmetric solutions.

\subsection{M-theory on $T^{1,1}$}
As we know M-theory compactification on a spatial 2-torus, $T^{0,2}$,
in the limit
of shrinking  torus leads to type IIB string theory.
Similarly,
M-theory compactification on a Lorentzian torus, $T^{1,1}$, in the limit
of torus shrinking to zero size leads to type IIB$^*$ strings
\cite{hull}. So there
 is way to expect
that the Dpp-wave is related to a solution of M-theory. On the other
hand, one can also do
the oxidation of the type IIA$^*$ string solution \eqn{funstr} to eleven
dimensions which will give us following $(9+2)$-dimensional M$^*$ theory
solution \cite{hull}
\bea\label{msol}
&&g_{(11)}:= W^{-{2\over3}}\left\{- (dx^{+})^2+(dx^{-})^2-dz^2\right\}
+W^{1\ov3} \sum_{a=1}^8(dx^a)^2\br
&& F_{(4)}\equiv dC_{(3)}=W^{-2}(\partial_a W) dx^adx^{+}dx^{-}dz +
2\sqrt{2}\mu
(dx^1dx^2dx^3dx^4+dx^5dx^6dx^7dx^8)\br
\eea
where $W$ is as in \eqn{funstr}. The coordinates $x^{+}, z$ are two time
like directions in M$^*$ theory.  It is useful to mantion that the
low energy supergravity action for M$^*$ theory of which eq.\eqn{msol}
is a background solution, has negative
kinetic terms for 3-form potential $C_{(3)}$ \cite{hull}.
 The solution \eqn{msol} has both electric
as well as magnetic type fluxes, hence it can be described as a bound
state of  2-branes and 5-branes of M$^*$ theory.

\section{Conclusions}
We have first shown that the Dpp-wave \eqn{newwave} is a
solution of the IIB* string theory which can be obtained by taking
pp-wave limits of the corresponding de Sitter solution.
These Dpp-waves are of a new kind in that they have the
properties
$$R_{--}<0$$
and entries of the matrix $ A_{ab}$ are strictly positive. These values
are  just opposite in sign compared to those of Hpp-waves.
Then we have shown that starting from Hpp-wave of IIB string
theory and by implementing a
time-like and a space-like T-duality in succession, we
obtain nothing but the Dpp-wave background.
This is confirmation of the fact
that the Hull's time-like duality map works for the case of Hpp-waves.
It will be interesting to study time-like duality maps for other pp-wave
backgrounds of string theory.

This study also provides us with a classfication of plane waves. Precisely
we have got three types of plane-wave space-times; viz, the {\it
Hpp-waves}
\cite{blau}

i) $R_{--}>0$ with $A_{ab}<0, ~Tr A_{ab}\ne 0 $,

the {\it  plane-waves without matter} \cite{horowitz}

ii) $R_{--}=0$ with $Tr A_{ab}=0$

 and the IIB$^*$ {\it Dpp-waves} with

iii) $R_{--}<0$ with $A_{ab}>0,~Tr A_{ab}\ne 0$.

It would be  worthwhile to study the BMN-correspondence for the
case of
Dpp-waves. Though at the first place it is not clear what is the
exact nature of $dS/CFT$
correspondence for de Sitter backgrounds \cite{witten}, which are
natural solutions of type II$^*$
string theories.

\leftline{\bf Acknowledgements}
{I am grateful to the IIT, Guwahati for hiring me. I am thankful to Rajesh
Gopakumar for useful discussion.
I would also like to thank the organisors of "National Workshop
on String Theory" at Indian Institute of Technology, Kanpur
for warm hospitality where this work was partly presented.
}

\vskip .5cm
\appendix{
\section{\underline {\bf Action in  type II$^*$ supergravity }}

The type IIA$^*$ effective action with $(1,9)$ spacetime signature is
given by
\cite{hull}
\begin{eqnarray}
S&=&\int \bigg[ e^{-2\f}\left\{
R~^{\ast}1+4d\f~^{\ast} d\f -
{1\over2} H_{(3)} ~^{\ast} H_{(3)}\right\}
+{1\over2} F_{(2)} ~^{\ast} F_{(2)} +{1\over2} F_{(4)}
~^{\ast} F_{(4)} \br
&& +{1\ov2} d C_{(3)} d C_{(3)}  B_{(2)} +\cdots\bigg]\ ,
\lll{IIa}
\end{eqnarray}
where all Ramond-Ramond kinetic terms are ghost-like. The field
strengths are defined through $F_{(n)}=dC_{n-1}+\cdots$. It is the same
thing for type
IIB$^*$ string theory
\begin{eqnarray}
S&=&\int \bigg[ e^{-2\f}\left\{
R~^{\ast}1+4d\f~^{\ast} d\f -
{1\over2} H_{(3)} ~^{\ast} H_{(3)}\right\}
+{1\over2} F_{(1)} ~^{\ast} F_{(1)} +{1\over2} F_{(5)}
~^{\ast} F_{(5)} +\cdots\bigg] \ ,\br
\lll{IIb}
\end{eqnarray}
for which equations of motion are supplemented with the
constraint equation that five-form field
strength satisfies self-dual  equation
$F_{(5)}=^* F_{(5)}$, where ${}^*$ is the Hodge-dual in ten dimensions.

For a time-like T-duality in the $X^0$ direction the relations between
dual \cite{hull}
backgrounds are
\bea\label{tdual}
&&\tilde g_{00}={1\over g_{00}}\br
&& \tilde g_{0\a}={B_{0\a}\over g_{00}}\br
&& \tilde B_{0\a}={g_{0\a}\over g_{00}}\br
&&\tilde g_{\a\b}=g_{\a\b} -{g_{0\a}g_{0\b}-B_{0\a}B_{0\b}\over
g_{00}} \br
&&\tilde B_{\a\b}=B_{\a\b} -{g_{0\a}B_{0\b}-B_{0\a}g_{0\b}\over
g_{00}} \br
&& \tilde{\f}=\f -{1\over2}log|g_{00}|
\eea
along with the RR-fields strengths $F_n$ of the type IIA (IIB)
theory related  to the field strengths $\tilde F_{n\pm 1}$ of the
dual type IIB$^*$ (IIA$^*$) theory as
$$ \tilde F_{0\m_1\cdots\m_n}=
F_{\m_{1}\cdots\m_n}, ~~~~
\tilde F_{\m_{1}\cdots\m_n}=-
F_{0\m_{1}\cdots\m_n} $$
for any $\m_i\ne0$.

}

\end{document}